\documentstyle[a4wide,epsf,11pt,titlepage]{article}

\newcounter{nref}
\newcommand{\bbib}{%
  \renewcommand{\refname}{\large\bf References}%
  \begin{thebibliography}{99}%
  \setcounter{enumiv}{\thenref}}
\newcommand{\ebib}{%
  \end{thebibliography}%
  \setcounter{nref}{\arabic{enumiv}}}
\newcommand{\head}[3]{%
  \setcounter{nref}{0}%
  \thispagestyle{empty}%
  \section*{\LARGE\bf #1}%
  \stepcounter{section}%
  \addcontentsline{toc}{section}{#1}%
  \large\itshape%
  #2\\\vspace{0.1pt}\\%
  #3%
  \normalsize\upshape%
  \bigskip}

\begin{document}


\head{Statistical mechanics of 2D turbulence with a prior vorticity distribution}
     {P.H. \ Chavanis} {Laboratoire de Physique Th\'eorique\\
     Universit\'e Paul Sabatier \\ 118, route de Narbonne\\ 31062 Toulouse,
     France}

\subsection*{Abstract}

We adapt the formalism of the statistical theory of 2D turbulence in
the case where the Casimir constraints are replaced by the
specification of a prior vorticity distribution. A new relaxation
equation is obtained for the evolution of the coarse-grained
vorticity.  It can be used as a thermodynamical parametrization of
forced 2D turbulence (determined by the prior), or as a numerical
algorithm to construct arbitrary nonlinearly dynamically stable
stationary solutions of the 2D Euler equation.

\vskip0.5cm

Two-dimensional incompressible flows with high Reynolds numbers are
described by the 2D Euler equations
\begin{equation}
{\partial \omega\over\partial t}+{\bf u}\cdot \nabla \omega=0,\qquad \omega=-\Delta\psi,\qquad {\bf u}=-{\bf z}\times\nabla\psi,
\label{t1}
\end{equation}
where $\omega$ is the vorticity and $\psi$ the streamfunction.  The 2D
Euler equations are known to develop a complicated mixing process
which ultimately leads to the emergence of a large-scale coherent
structure, typically a jet or a vortex \cite{houches}. Jovian
atmosphere shows a wide diversity of structures: Jupiter's great red
spot, white ovals, brown barges,... One question of fundamental
interest is to understand and predict the structure and the stability
of these equilibrium states. To that purpose, Miller \cite{miller} and
Robert \& Sommeria
\cite{rs} have proposed a statistical mechanics of the 2D Euler
equation. The idea is to replace the deterministic description of the
flow $\omega({\bf r},t)$ by a probabilistic description where
$\rho({\bf r},\sigma,t)$ gives the density probability of finding the
vorticity level $\omega=\sigma$ in ${\bf r}$ at time $t$. The observed
(coarse-grained) vorticity field is then expressed as
$\overline{\omega}({\bf r},t)=\int \rho\sigma d\sigma$. To apply the
statistical theory, one must first specify the constraints attached to
the 2D Euler equation. The circulation $\Gamma=\int
\overline{\omega}d{\bf r}$ and the energy $E={1\over 2}\int
\overline{\omega}\psi d{\bf r}$ will be called {\it robust
constraints} because they can be expressed in terms of the
coarse-grained field $\overline{\omega}$ (the energy of the
fluctuations can be neglected). These integrals can be
calculated at any time from the coarse-grained field
$\overline{\omega}({\bf r},t)$ and they are conserved by the dynamics. By contrast, the Casimir invariants $I_{f}=\int
\overline{f(\omega)}d{\bf r}$, or equivalently the fine-grained
moments of the vorticity $\Gamma_{n>1}^{f.g.}=\int
\overline{\omega^{n}}d{\bf r}=\int \rho\sigma^{n}d\sigma d{\bf r}$,
will be called {\it fragile constraints} because they must be
expressed in terms of the fine-grained vorticity. Indeed, the moments of the
coarse-grained vorticity $\Gamma_{n>1}^{c.g}=\int
\overline{\omega}^{n}d{\bf r}$ are not conserved since
$\overline{\omega^{n}}\neq \overline{\omega}^{n}$ (part of the
coarse-grained moments goes into fine-grained
fluctuations). Therefore, the moments $\Gamma_{n>1}^{f.g.}$ must be
calculated from the fine-grained field $\omega({\bf r},t)$ or from the
initial conditions, i.e. before the vorticity has mixed. Since we
often do not know the initial conditions nor the fine-grained field,
the Casimir invariants often appear as ``hidden constraints''
\cite{super}. 

The statistical theory of Miller-Robert-Sommeria is based on two
assumptions: (i) it is assumed that we know the initial conditions (or
equivalently the value of all the Casimirs) in detail (ii) it is
assumed that mixing is efficient and that the evolution is ergodic so
that the system will reach at equilibrium the most probable (most
mixed) state. Within these assumptions\footnote{Some attempts have
been proposed to go beyond the assumptions of the statistical
theory. For example, Chavanis \& Sommeria \cite{jfm1} consider a {\it
strong mixing limit} in which only the first moments of the vorticity
are relevant instead of the whole set of Casimirs. On the other hand,
Chavanis \& Sommeria \cite{jfm2} introduce the concept of {\it maximum
entropy bubbles} (or restricted equilibrium states) in order to account
for situations where the evolution of the flow is not ergodic in the
whole available domain but only in a subdomain.}, the statistical
equilibrium state of the 2D Euler equation is obtained by maximizing
the mixing entropy
\begin{equation}
\label{t2}
S\lbrack \rho\rbrack=-\int \rho\ln\rho \ d{\bf r}d\sigma,
\end{equation}  
at fixed energy $E$ and circulation $\Gamma$ (robust constraints) and fixed fine-grained moments $\Gamma_{n>1}^{f.g.}$ (fragile constraints). This optimization principle is solved by introducing Lagrange multipliers, writing the first order variations as
\begin{equation}
\label{t3}
\delta S-\beta\delta E-\alpha\delta\Gamma-\sum_{n>1}\alpha_{n}\delta\Gamma^{f.g.}_{n}=0.
\end{equation}

In the approach of Miller-Robert-Sommeria, it is assumed that the
system is strictly described by the 2D Euler equation so that the
conservation of all the Casimirs has to be taken into account.
However, in geophysical situations, the flows are forced and
dissipated at small scales (due to convection in the jovian
atmosphere) so that the conservation of the Casimirs is
destroyed. Ellis {\it et al.} \cite{ellis} have proposed to treat
these situations by fixing the conjugate variables $\alpha_{n>1}$
instead of the fragile moments $\Gamma_{n>1}^{f.g.}$. If we view the
vorticity levels as species of particles, this is similar to fixing
the chemical potentials instead of the total number of particles in
each species. Therefore, the idea is to treat the fragile constraints
{\it canonically}, whereas the robust constraints are still treated
{\it microcanonically}. This point of view has been further developed
in Chavanis \cite{physicaD}. The relevant thermodynamical potential is
obtained from the mixing entropy (\ref{t2}) by using a Legendre
transform with respect to the fragile constraints \cite{physicaD}:
\begin{equation}
\label{t4}
S_{\chi}= S-\sum_{n>1}\alpha_{n}\ \Gamma^{f.g.}_{n}.
\end{equation}
Expliciting the fine-grained moments, we obtain the {\it relative entropy} 
\begin{equation}
\label{t5}
S_{\chi}\lbrack \rho\rbrack=-\int \rho\ \ln\biggl\lbrack {\rho\over \chi(\sigma)}\biggr\rbrack \ d{\bf r}d\sigma,
\end{equation}
where we have defined the {\it prior vorticity distribution}
\begin{equation}
\label{t6}
\chi(\sigma)\equiv{\rm exp}\biggl\lbrace -\sum_{n>1}\alpha_{n}\sigma^{n}\biggr\rbrace.
\end{equation}
We shall assume that this function is {\it imposed} by the small-scale
forcing. Assuming ergodicity, the statistical equilibrium state is now
obtained by maximizing the relative entropy $S_{\chi}$ at fixed energy
$E$ and circulation $\Gamma$ (no other constraints). The conservation
of the Casimirs has been replaced by the specification of the prior
$\chi(\sigma)$. Writing $\delta S_{\chi}-\beta\delta
E-\alpha\delta\Gamma=0$, and accounting for the normalization
condition $\int\rho d\sigma=1$, we get the Gibbs state
\begin{equation}
\label{t7}
\rho({\bf r},\sigma)={1\over Z({\bf r})}\chi(\sigma)  e^{-(\beta\psi+\alpha)\sigma}\quad {\rm with}\quad  Z=\int_{-\infty}^{+\infty} \chi(\sigma)  e^{-(\beta\psi+\alpha)\sigma}d\sigma.
\end{equation}
This is the product of a universal Boltzmann factor by a non-universal function $\chi(\sigma)$ fixed by the forcing. The coarse-grained vorticity is given by
\begin{equation}
\label{t8}
\overline{\omega}={\int \chi(\sigma)\sigma e^{-(\beta\psi+\alpha)\sigma} d\sigma\over \int \chi(\sigma) e^{-(\beta\psi+\alpha)\sigma} d\sigma}=F(\beta\psi+\alpha)\quad {\rm with}\quad F(\Phi)=-(\ln\hat\chi)'(\Phi),
\end{equation}
where $\hat{\chi}(\Phi)=\int_{-\infty}^{+\infty}\chi(\sigma)
e^{-\sigma\Phi}d\sigma$. It is easy to show that
$F'(\Phi)=-\omega_{2}(\Phi)\le 0$, where
$\omega_{2}=\overline{\omega^{2}}-\overline{\omega}^{2}\ge 0$ is the
local centered variance of the vorticity, so that $F$ is a decreasing
function \cite{sw}.  Therefore, the statistical theory predicts that
the coarse-grained vorticity $\overline{\omega}=f(\psi)$ is a {\it
stationary solution} of the 2D Euler equation and that the
$\overline{\omega}-\psi$ relationship is a {\it monotonic} function
which is increasing at {negative temperatures} $\beta<0$ and
decreasing at positive temperatures $\beta>0$ since
$\overline{\omega}'(\psi)=-\beta\omega_{2}$.  We also note that the
most probable vorticity $\langle \sigma\rangle({\bf r})$ of the
distribution (\ref{t7}) is given by
\cite{nic}:
\begin{equation}
\label{t9}
\langle \sigma\rangle=\lbrack (\ln\chi)'\rbrack^{-1}(\beta\psi+\alpha),
\end{equation}
provided $(\ln\chi)''(\langle\sigma\rangle)<0$.  This is also a
stationary solution of the 2D Euler equation which usually differs
from the average value $\overline{\omega}({\bf r})$ of the
distribution (\ref{t7}) except when $\chi(\sigma)$ is gaussian. We
note that the $\overline{\omega}-\psi$ relationship predicted by the
statistical theory can take a wide diversity of forms
(non-Boltzmannian) depending on the prior $\chi(\sigma)$. The
coarse-grained vorticity (\ref{t8}) can be viewed as a sort of {\it
superstatistics} as it is expressed as a superposition of Boltzmann
factors (on the fine-grained scale) weighted by a non-universal
function $\chi(\sigma)$
\cite{super}. Furthermore, the coarse-grained vorticity (\ref{t8})
maximizes a generalized entropy (in $\overline{\omega}$-space) of the
form \cite{pre}:
\begin{equation}
\label{t10}
S\lbrack \overline{\omega}\rbrack =-\int C(\overline{\omega})  d{\bf r},
\end{equation}
at fixed circulation and energy (robust constraints). Writing $\delta
S-\beta\delta E-\alpha\delta \Gamma=0$ leading to
$C'(\overline{\omega})=-\beta\psi-\alpha$ and
$\overline{\omega}'(\psi)=-\beta/C''(\overline{\omega})$, and
comparing with Eq. (\ref{t8}), we find that $C$ is a convex function
($C''>0$) determined by the prior $\chi(\sigma)$ encoding the
small-scale forcing according to the relation \cite{super}:
\begin{equation}
\label{t11} C(\overline{\omega})=-\int^{\overline{\omega}}F^{-1}(x)dx=-\int^{\overline{\omega}}\lbrack (\ln {\hat \chi})'\rbrack^{-1}(-x)dx.
\end{equation}
The preceding relations are also valid in the approach of
Miller-Robert-Sommeria except that $\chi(\sigma)$ is determined {\it a
posteriori} from the initial conditions by relating the Lagrange
multipliers $\alpha_{n>1}$ to the Casimir constraints
$\Gamma^{f.g.}_{n>1}$. In this case of freely evolving flows, the
generalized entropy (\ref{t10}) depends on the initial conditions,
while in the case of forced flows considered here, it is intrinsically
fixed by the prior vorticity distribution.

In that context, it is possible to propose a thermodynamical
parameterization of 2D forced turbulence in the form of a relaxation
equation that conserves circulation and energy (robust constraints)
and that increases the generalized entropy (\ref{t10}) fixed by the
prior $\chi(\sigma)$. This equation can be obtained from a generalized Maximum
Entropy Production (MEP) principle in $\overline{\omega}$-space \cite{pre} by
writing the coarse-grained 2D Euler equation in the form
$D\overline{\omega}/Dt=-\nabla\cdot \overline{\tilde\omega\tilde {\bf
u}}=-\nabla\cdot {\bf J}$ and determining the optimal current ${\bf
J}$ which maximizes the rate of entropy production $\dot S=-\int
C''(\overline{\omega}){\bf J}\cdot \nabla\overline{\omega}d{\bf r}$ at
fixed energy $\dot E=\int {\bf J}\cdot \nabla\psi d{\bf r}=0$, assuming
that the energy of fluctuations ${\bf J}^{2}/2\overline{\omega}$ is
bounded. According to this principle, we find that the coarse-grained
vorticity evolves according to \cite{pre,physicaD}:
\begin{equation}
{\partial \overline{\omega}\over\partial t}+{\bf u}\cdot \nabla
\overline{\omega}=\nabla\cdot \biggl \lbrace D
\biggl\lbrack \nabla\overline{\omega}+{\beta(t)\over
C''(\overline{\omega})}\nabla\psi\biggr\rbrack\biggr\rbrace ,\qquad
\overline{\omega}=-\Delta\psi,
\label{t12}
\end{equation}
\begin{equation}
\beta(t)=-{\int D\nabla\overline{\omega}\cdot\nabla\psi d^{2}{\bf r}\over \int D{(\nabla\psi)^{2}\over C''(\overline{\omega})}d^{2}{\bf r}}, \qquad D\propto \omega_{2}^{1/2}={1\over \sqrt{C''(\overline{\omega})}},
\label{t13}
\end{equation} 
where $\beta(t)$ is a Lagrange multiplier enforcing the energy
constraint $\dot E=0$ at any time. These equations increase the
entropy ($H$-theorem $\dot S\ge 0$) provided that $D>0$, until the
equilibrium state (\ref{t8}) is reached. The diffusion coefficient $D$
is not determined by the MEP but it can be obtained from a Taylor's
type argument leading to expression (13)-b \cite{physicaD}. This
diffusion coefficient, related to the strength of the fluctuations,
can ``freeze'' the relaxation in a sub-region of space (``bubble'')
and account for {\it incomplete relaxation} and lack of ergodicity
\cite{rr,csr}. The relaxation equation (\ref{t12}) belongs to the
class of generalized Fokker-Planck equations introduced in Chavanis
\cite{pre}. This relaxation equation conserves only the robust
constraints (circulation and energy) and increases the generalized
entropy (\ref{t11}) fixed by the prior vorticity distribution
$\chi(\sigma)$.  It differs from the relaxation equations proposed by
Robert \& Sommeria \cite{rsmepp} for freely evolving flows which
conserve all the constraints of the 2D Euler equation (including all
the Casimirs) and increase the mixing entropy (\ref{t2}). In
Eqs. (\ref{t12})-(\ref{t13}), the specification of the prior
$\chi(\sigma)$ (determined by the small-scale forcing) replaces the
specification of the Casimirs (determined by the initial conditions).
However, in both models, the robust constraints $E$ and $\Gamma$ are
treated microcanonically (i.e. they are rigorously
conserved). Furthermore, in the two-levels case $\omega\in
\lbrace\sigma_{0},\sigma_{1}\rbrace$, the two approaches are formally
equivalent and they amount to maximizing a generalized entropy
(\ref{t10}) similar to the Fermi-Dirac entropy at fixed circulation
and energy \cite{csr}.  In the viewpoint of
Miller-Robert-Sommeria, this entropy describes the free merging of a
system with two levels of vorticity $\sigma_{0}$ and $\sigma_{1}$
while in the other viewpoint, it describes the evolution of a forced
system where the forcing has two intense peaks described by the prior
$\chi(\sigma)=\chi_{0}\delta(\sigma-\sigma_{0})+\chi_{1}\delta(\sigma-\sigma_{1})$ \cite{physicaD}.

The relaxation equations (\ref{t12})-(\ref{t13}) can also be used as a
{\it numerical algorithm} to construct stable stationary solutions of
the 2D Euler equation. Indeed, Ellis {\it et al.} \cite{ellis} have
shown that the maximization of a functional of the form (\ref{t10}) at
fixed energy and circulation determines a stationary solution of the
2D Euler equation of the form $\omega=f(\psi)$, where $f$ is
monotonic, which is nonlinearly dynamically stable. Since the
stationary solution of Eqs.  (\ref{t12})-(\ref{t13}) maximizes $S$ at
fixed $E$ and $\Gamma$ (by construction), this steady solution of the
relaxation equations is also a nonlinearly dynamically stable
stationary solution of the 2D Euler equations (\ref{t1}). Thus, by
changing the convex function $C(\omega)$ in Eq. (\ref{t12}), we can
numerically construct a wide diversity of stable solutions of the 2D Euler
equations. This is a potentially interesting procedure because it is
usually difficult to solve the differential equation
$-\Delta\psi=f(\psi)$ directly and be sure that the solution is
(nonlinearly) dynamically stable.  These nonlinearly stable steady
states can be an alternative to the statistical equilibrium state in
case of incomplete relaxation, when the system has not mixed
efficiently (non-ergodicity) so that the statistical prediction fails.
In case of incomplete relaxation we cannot {\it predict} the
equilibrium state but we can try to {\it reproduce} it a posteriori.

Finally, we have proposed in \cite{pre} to develop a
phenomenological/effective statistical theory of 2D turbulence to deal
with complex situations. The idea is that some types of entropic
functional $S[\overline{\omega}]$ (in $\overline{\omega}$-space) may
be more appropriate than others to describe a given physical
situation. For example, the enstrophy functional turns out to be
relevant in certain oceanic situations \cite{verron} and the
Fermi-Dirac type entropy in jovian flows \cite{bs,sw}. Certainly,
other functionals of the same ``class'' would work as well for these
systems. In addition, other classes of functionals
$S[\overline{\omega}]$ may be relevant in other
circumstances. Therefore, as a simple and practical procedure to
describe a given system, we propose to pick a functional
$S[\overline{\omega}]$ in the ``class of equivalence'' appropriate to
that system and use it in the parameterization
(\ref{t12})-(\ref{t13}). We can thus describe the time evolution of
the system on the coarse-grained scale. This approach is not
completely predictive because we need to know in advance which type of
entropy $S[\overline{\omega}]$ describes best such and such
situation. In practice, it must be determined by trying and errors
(e.g. by comparing with oceanic data). But once a specific entropy has
been found for a physical situation, we can work with it for different
initial conditions specified by the robust constraints $E$ and
$\Gamma$ (the effect of the Casimirs is reported in the chosen form of
entropy $S[\overline{\omega}]$).  The idea is that the entropy $S$
remains the same while $E$ and $\Gamma$ are changed. The problem is
rich and non-trivial even if $S$ has been fixed because bifurcations
can occur depending on the control parameters $E$, $\Gamma$.  This
heuristic approach can be viewed as a simple attempt to account for
the influence of the Casimirs while leaving the problem tractable. We
use the fact that the Casimirs lead to {\it non-standard}
(i.e. non-Boltzmannian) $\overline{\omega}-\psi$ relationships at
equilibrium which are associated with non-standard forms of entropy
$S[\overline{\omega}]$ in $\overline{\omega}$-space. We propose to
{\it fix} the $S$-functional depending on the situation. We do not try
to {\it predict} its form, but rather to {\it adjust} it to the
situation contemplated. This is based on the belief that some
functionals $S[\overline{\omega}]$ are more relevant than others for a
given system. Whether this is the case or not remains to be
established. All the ideas presented here can be generalized to the
case of quasi-geostrophic or shallow-water equations \cite{sw}.



\bbib
\bibitem{houches}  {\small P.H. Chavanis, {\it Statistical mechanics of two-dimensional vortices and stellar systems}, in {\it Dynamics and thermodynamics of systems with long range interactions}, edited by Dauxois, T., Ruffo, S., Arimondo, E. \& Wilkens, M. Lecture Notes in Physics, Springer (2002); [cond-mat/0212223]. }
\bibitem{miller} J.~Miller, Phys. Rev. Lett.
    {\bf 65} (1990) 2137.
\bibitem{rs} R.~Robert and J.~Sommeria, J. Fluid Mech.
    {\bf 229} (1991) 291. 
\bibitem{super} P.H.~Chavanis, {\it Coarse-grained distributions and superstatistics} [cond-mat/0409511]
\bibitem{jfm1} P.H.~Chavanis and J.~Sommeria, J. Fluid Mech.
    {\bf 314} (1996) 267. 
\bibitem{jfm2} P.H.~Chavanis and J.~Sommeria, J. Fluid Mech.
    {\bf 356} (1998) 259.
\bibitem{ellis} R.~Ellis, K.~Haven and B.~Turkington, Nonlinearity
    {\bf 15} (2002) 239. 
\bibitem{physicaD} P.H.~Chavanis, Physica D
    {\bf 200} (2005) 257. 
\bibitem{sw} P.H.~Chavanis and J.~Sommeria, Phys. Rev. E
    {\bf 65} (2002)  026302. 
\bibitem{nic} N. Leprovost, B. Dubrulle and P.H.~Chavanis, {\it Dynamics and thermodynamics of axisymmetric flows: I. Theory} [physics/0505084]
\bibitem{pre} P.H.~Chavanis, Phys. Rev. E
    {\bf 68} (2003) 036108. 
\bibitem{rr} R.~Robert and C.~Rosier, J. Stat. Phys.
    {\bf 86} (1997) 481. 
\bibitem{csr} P.H.~Chavanis, J.~Sommeria and R.~Robert, Astrophys. J.    {\bf 471} (1996) 385. 
\bibitem{rsmepp} R.~Robert and J.~Sommeria, Phys. Rev. Lett.
    {\bf 69} (1992) 2776. 
\bibitem{verron} E.~Kazantsev, J.~Sommeria and J.~Verron, J. Phys. Oceanogr.
    {\bf 28} (1998) 1017. 
\bibitem{bs} F.~Bouchet and J.~Sommeria, J. Fluid Mech.
    {\bf 464} (2002) 165. 
\ebib
\end{document}